\documentclass[aps,twocolumn,prb,showpacs,superscriptaddress,preprintnumbers,amsmath,amssymb]{revtex4-1}

\usepackage{graphicx}% Include figure files
\usepackage{dcolumn}% Align table columns on decimal point
\usepackage{bm}% bold math
\usepackage{color}
\usepackage[latin1]{inputenc}
\usepackage[hang]{footmisc}
\usepackage{wasysym}
\usepackage{verbatim}
\begin{document}

% once the note is approved for conferences, remove the following lines
%\begin{widetext}
%\center{INTERNAL DOCUMENT -- NOT FOR PUBLIC DISTRIBUTION}
%\end{widetext}

\title{On the Origin of Non-Saturating Linear Magnetoresistivity}

%\\ \vspace*{2.0cm}}
% throughout the internal review, the note will be authored by individuals
% who contributed to the paper
\author{Ferdinand Kisslinger}
%\thanks{}
\affiliation{Lehrstuhl f\"ur Angewandte Physik,
Friedrich-Alexander-Universit\"at Erlangen-N\"urnberg (FAU), Staudtstr.\ 7,
91058 Erlangen, Germany}
%\affiliation{Dresden High Magnetic Field Laboratory,
%Helmholtz-Zentrum Dresden-Rossendorf, Bautzner Landstraße 400,
%01328 Dresden, Germany}

\author{Christian Ott}
%\thanks{}
\affiliation{Lehrstuhl f\"ur Angewandte Physik,
Friedrich-Alexander-Universit\"at Erlangen-N\"urnberg (FAU), Staudtstr.\ 7,
91058 Erlangen, Germany}

\author{Heiko B. Weber}

\email[]{heiko.weber@fau.de}

%\altaffiliation{Interdisziplin\"ares Zentrum f\"ur Molekulare
%Materialien, Friedrich-Alexander-Universit\"at Erlangen-N\"urnberg}

%\affiliation{Lehrstuhl f\"ur Angewandte Physik,
%Friedrich-Alexander-Universit\"at Erlangen-N\"urnberg (FAU), Staudtstr.\ 7,
%91058 Erlangen, Germany}

%\homepage[http://www.lap.physik.fau.de]
%\thanks{}

% Also, during the review process,
% include the header:

%\begin{widetext}
%Version: 01   Send comments to stefan.ballmann@physik.uni-erlangen.de \\
%Author(s): WS     June, 23rd 2010
%\end{widetext}

% once the note is approved for conferences, remove the above header

\date{\today}

\begin{abstract}
% remove the space for publication
%\vspace*{3.0cm}
  The observation of non-saturating classical linear magnetoresistivity has been an enigmatic phenomenon in solid state physics. We present a study of a two-dimensional ohmic conductor, including local Hall effect and a self-consistent consideration of the environment. An equivalent-circuit scheme delivers a simple and convincing argument why the magnetoresistivity is linear in strong magnetic field, provided that current and biasing electric field are misaligned by a nonlocal mechanism. A finite-element model of a two-dimensional conductor is suited to display the situations that create such deviating currents. Besides edge effects next to electrodes, charge carrier density fluctuations are efficiently generating this effect. However, mobility fluctuations that have frequently been related to linear magnetoresistivity are barely relevant. Despite its rare observation, linear magnetoresitivity is rather the rule than the exception in a regime of low charge carrier densities, misaligned current pathways and strong magnetic field.
% remove this for publication
%\vspace*{5.0cm}
%\centerline{\em Preliminary Manuscript}
\end{abstract}

% activate the following line for publication
%\pacs{85.65.+h, 72.10.Di, 73.23.-b, 73.40.Gk, 63.22.-m, 81.07.Nb}

\maketitle

\section{Introduction}

The classical magnetoresistivity $\rho (B)$ (mr) of a homogeneous conductor vanishes in the simplest models, but under realistic assumptions it is quadratic in magnetic field $B$ and starts to saturate when $\mu B$ exceeds unity ($\mu$ is the charge carrier mobility)\cite{popovic,Pippard}. However, since the early days of solid state physics, counterexamples are known, for which the mr is strictly linear, without saturation \cite{Kapitza,Pippard}. This phenomenon remains a barely resolved enigma of solid state physics and has lead to significant confusion. It has been reported in experiments on three-dimensional materials as different as common metals \cite{taub}, semimetals \cite{McClure,Xu,Husmann,Lee,Lee15}, and semiconductors \cite{Delmo,Porter,Kozlova,Hu2008}. As magnetoresistivity is essentially a two-dimensional phenomenon, it occurs also in the novel 2D materials classes of graphene-derivated materials \cite{Friedman,Kisslinger, Zhi} and topological insulators \cite{WangPRL2012,WangSciRep,Assaf,Narayanan,Wang2014}. The enigma is, which generic mechanism creates such a simple phenomenology (strictly linear mr, no saturation) that is seemingly insensitive to the substantial differences provided by the broad range of materials that have only a finite conductivity in common. 

There are models which attempt to explain linear mr as a quantum phenomenon at the lowest Landau level \cite{Abrikosov,Abrikosov2000}, or near charge neutrality \cite{gornyi}. Another approach has been presented in 2003 by Parish and Littlewood (PL), which is essentially a finite-element analysis of the Magnetoresistance (MR, in contrast to the magnetoresistivity mr) of a finite 2D conductor, including Hall effect and Kirchhoff rules. This model guides the way to the correct understanding, but remains incomplete and caused significant confusion. We will critically discuss it in more detail below (see section \ref{sec:CriticalReview}).

\begin{figure*}[hbtp!]
    \centering
        \includegraphics[width=178mm]{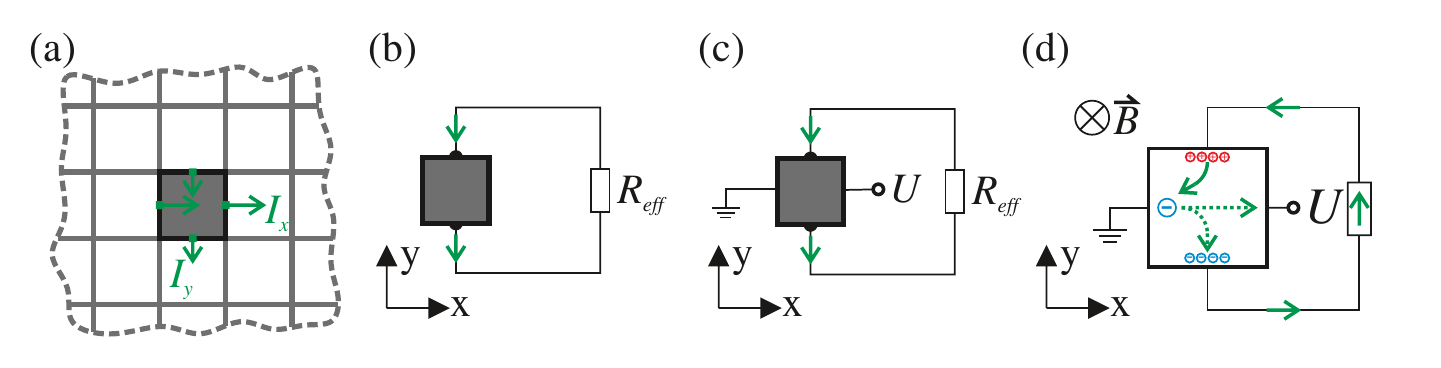}
        \caption{(Color online): 
	(a) Homogeneous conductor subdivided in small square tiles. Arbitrary tile (grey), far away from electrode banks and boundaries, where translational symmetry is fulfilled. (b-d) Stepwise design of an equivalent circuit for such a tile. (b) Current conservation in $y$ direction is guaranteed \textit{via} a feedback resistor $R_{eff}(B)$. (c) A voltage source is added to drive a current. (d) Schematic of the electron current flow in the equivalent circuit. The currents $I_x$ and $I_y$ are independently conserved by the circuit, although they are mixed in the presence of magnetic field.} \label{fig:1}
\end{figure*}

\section{The generating mechanism, or why linear?} \label{sec:generating_mechanism}

The resistivity tensor in a homogeneous isotropic 2D material in a perpendicular magnetic field $B$ reads 
\begin{align}
	\begin{pmatrix}
		E_x \\
		E_y
	\end{pmatrix} =
	\begin{pmatrix}
		\rho_0 & B/ne \\
		-B/ne & \rho_0
	\end{pmatrix} \cdot
	\begin{pmatrix}
		j_x \\
		j_y
	\end{pmatrix} \quad \text{,}
\end{align}
with $\boldsymbol{E}$ the electric field, $\boldsymbol{j}$ the current density, $\rho_0=1/ne\mu$ the Drude resistivity and $n$ the charge carrier density. For zero magnetic field $B$, current and electric field are collinear. For strong magnetic fields, defined by $\mu B \gg 1$, these quantities approach to a perpendicular configuration, which leads to a reshaped potential landscape. As a first approach to the problem, we treat this limit such that we suppress the diagonal terms $\rho_0 $ entirely. Then, the equations read
\begin{align}
E_x & = B/ne j_y \\
E_y & = -B/ne j_x
\end{align}

 In this limit, currents flow along equipotential lines, and further increase of $B$ does not affect the direction of currents, neither it modifies the potential landscape further (under voltage biased conditions). This is a well-known phenomenon that can be studied either analytically \cite{Lippmann} or using finite element analyses \cite{PL1,PL2} and will be discussed in detail later in the manuscript. When $E_x$ and $E_y$ are frozen, these equations leave no other solution than $j \propto 1/B$. This simple scaling argument is the origin of linear mr. Note that in addition to the local resistance tensor that has no mr, a non-local property of the environment ($B$-independent $\boldsymbol{E}$) is important in order to explain linear mr.

%At any position of such a potential map, the current ratio $j_y/j_x =: \tan{\alpha}$ can be defined, that depends on $B$ but approaches asymptotically to a constant for strong $B$. 

%In a real sample of finite size under voltage biased conditions, $E_x$ and $E_y$ can be read off a potential map, and finite size studies similar to \cite{Lippmann, PL1, PL2} can be undertaken either analytically or numerically. Extensive numerical studies are given below. One common observation in regularly shaped samples is that at very high magnetic fields, the potential maps asymptotically freezes, i.e. it do not change significantly when the magnetic field increases further. In specific geometries, deviations from this freezing can be observed (see SI). When the potential map, however, is frozen, this means that $E_x$ and $E_y$ do not depend on further increase of $B$. As a simple consequence, the products $B \cdot j_x$ as well as $B \cdot j_y$ become independent of $B$, or all current densities $j$ become inversely proportional to $B$. This is translated as linear magnetoresistivity.  
%At any point of the potential map, the current ratio $j_y/j_x =: \tan{\alpha}$ can be defined and is constant for strong $B$. Note that $\tan{\alpha (B)}= -\tan{\alpha (-B)}$. 

In order to link up with finite-element concepts, we focus on a single square tile in the middle of a homogeneous quasi-infinite conducting material with translational invariance (see Fig.\ref{fig:1}\,a). %The ingredients to the model are Hall effect, Kirchhoff rules and ohmic conduction. 
The relation between currents and voltages under the influence of Hall effect at the square tile is given by a resistance tensor:
\begin{align}
	\begin{pmatrix}
		U_x \\
		U_y
	\end{pmatrix} =
	\begin{pmatrix}
		\rho_0 & B/ne \\
		-B/ne & \rho_0
	\end{pmatrix} \cdot
	\begin{pmatrix}
		I_x \\
		I_y
	\end{pmatrix} \quad \text{,}
\end{align}
 We are interested in linear response conductivity, and assume without loss of generality a finite external voltage in $x$-direction that drives a current through the tile. Translational invariance guarantees that the current incoming from the left equals the current outgoing to the right, such that only one current $I_x$ (and, by the same argument, $I_y$) has to be considered (Fig.\ref{fig:1}\,a-d).

The model which we propose uses an equivalent circuit in which the top and bottom edge are shorted via an effective resistor $R_{eff}$ to ensure this current conservation in the calculation. The latter represents the effective resistance of the environment, summarizing many current paths that reach out in the conductive plane, with $R_y^\prime = -U_y/I_y = R_{eff}$. Solving this equivalent circuit one obtains the longitudinal resistance $R_x^\prime = U_x/I_x$ as

\begin{align}
	R_x^\prime = \rho_0 + \frac{(B/ne)^2}{\rho_0+R_{eff}}
\end{align}

As a special case, for $R_{eff} = \infty$ this formula contains the standard Hall result:  vanishing $I_y $, and a constant mr.
This system of equations, however, contains richer solutions with finite transversal currents $I_y$. We define $\tan{[\alpha(B)]}:=I_y/I_x$ as a measure of the direction of current with respect to the bias direction (= $x$-axis).

If one assumes $R_{eff}$ as constant (i.e. $B$ independent) then a quadratic mr $R_x^\prime \propto B^2$ would result in the strong-field limit. This, however, is obviously inconsistent: $R_{eff}$ represents the mr of the environment and has, therefore, the same $B$ dependence as the mr of the specific tile we have chosen. Self consistency  in the high-field limit ($\alpha=\alpha_\infty$) can be reached by choosing

%But asking for a self consistent solution, where the tile resistance has the same dependence on magnetic field as the environment, a factor linear in $B/ne$ appears. The self-consistent effecvbtive resistance is given by:

\begin{align}
	%\tan(\alpha_{\infty}) &\propto \frac{1}{R_{eff}} \cdot \frac{B}{ne}\label{eqn:6} \\
	R_{eff} &= \frac{1}{\tan(\alpha_\infty)} \cdot \frac{B}{ne}\label{eqn:6} \\
R_x^\prime &= U_x/I_x \approx \rho_0 + \tan(\alpha_\infty) \frac{B}{ne}, \label{eqn:4}
\end{align}
 
with $\tan(\alpha_\infty)$ appearing as the prefactor by comparison of coefficients (see SI). This is the only high-field solution, in which the tile`s effective resistance $R^\prime$ and its environment share the same $B$ dependence. This is the key argument why in the high-field limit the mr is linear. When rewriting (\ref{eqn:6}) as $\tan(\alpha_\infty) = \frac{1}{R_{eff}} \cdot \frac{B}{ne}$ it becomes obvious that $\alpha_\infty$ is determined by both, parameters of the tile (its charge density), as well as of the environment. It is therefore sensitive to the electrostatic landscape in which the tile is embedded. Note that within our simplifying model, $\alpha(B)$ is anti-symmetric with respect to $B$ and approaches in both field directions to asymptotic values $\alpha_{\infty}$ and $-\alpha_\infty$.

%It leaves, however, a prefactor $\tan{(\alpha)}$ undetermined, which can be either finite, or zero.  $\alpha$ indicates the direction of the current with respect to the applied voltage $U_x$, or, more generally, the angle between the incident current and the electric field.  $\alpha (B)$ depends on magnetic field and 

%Note that in the middle of a long Hall bar, the standard Hall condition ($\boldsymbol{I}$ parallel to the longitudinal voltage drop) leads to $\alpha = 0$, and vanishing mr results, in accordance with standard textbook results.

Following this argument, linear mr is the self-consistent solution of a simple conductor in the high-field limit and should therefore be rather the rule than the exception. The appearance of finite linear mr, however, requires an additional ingredient: a non-local mechanism that provides finite values of $\alpha_{\infty}$ and cants the current $\boldsymbol{I}$ with respect to the electric field bias $\boldsymbol{E}$ (in our model represented by a local voltage drop $U_x$). 

In the course of this manuscript, we focus on mechanisms that create such finite current distortion (CD) fields $\alpha_\infty (\boldsymbol{r})$ as a consequence of inhomogeneities. Among such we find the importance of macroscopic boundaries, addressed by PL, and bulk disorder, in particular local variations of the charge carrier density. 

\begin{figure*}
%\twocolcaption
  \includegraphics[width=178mm]{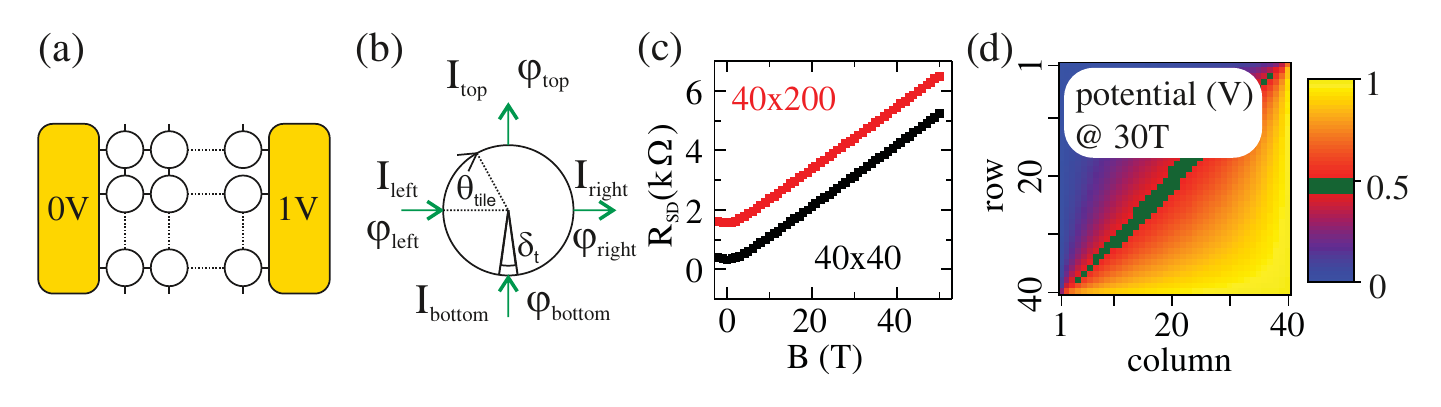}%
  \caption{(Color online): 
	(a) Model of a four-terminal network with equipotential electrode banks, very similar to \cite{PL1}. (b) Scheme of a four-terminal tile with currents $I$ (green arrows) and potentials $\varphi$ at the terminal positions. The opening angle of the terminal at angle $\theta_{tile}$ is given by $\delta_{t}$. (c) $R_{SD}(B)$ for homogeneous networks with $40 \times 40$ (lower black squares) and 40$\times$200 tiles (upper red squares). (d) Mean potential of each tile mapped for the 40$\times$40 network at a magnetic field of 30\,T ($\mu = 1$) with left electrode at 0\,V and right electrode at 1\,V. }\label{fig:2} 
\end{figure*}

\section{Methods} 

From now on, the paper essentially uses the same conceptual framework as PL \cite{PL1,PL2}, in particular a network of four-terminal (and three-terminal) tiles (Fig.\ref{fig:2}\,a) to model the 2D material.  Every little Hall tile in the network is treated as a classical conductor, the electrostatics of which obey Maxwell's equations. The classical transport in a circular tile (Fig. \ref{fig:2}\,b) is described by Ohm`s law $\boldsymbol{j}=\boldsymbol{\hat{\sigma}} \boldsymbol{E}$, with $\boldsymbol{j}$ the current density, $\boldsymbol{\hat{\sigma}}$ the conductivity tensor and $\boldsymbol{E} = - \boldsymbol{\nabla} \varphi$.The electrostatic potential $\varphi$ on the tile is calculated using the Laplace equation $\Delta \varphi = 0$, by expanding the solution in a Fourier series in the angle $\theta_{tile}$. The current density inserted into the tile is assumed to be constant over the whole opening angle $\delta_t$ of each terminal. Ohm`s law $- \boldsymbol{\hat{\sigma}} \cdot \boldsymbol{\nabla} \varphi  = \boldsymbol{j}$ can be solved, at the edge of the tile, by expanding the current density in a Fourier series and comparing coefficients of left and right hand side of the equation. Thus the impedance matrix $\boldsymbol{\hat{Z}}$ of the tile connecting the input currents $\boldsymbol{I}$ and the potentials $\boldsymbol{U}$ at the terminals is obtained (example for a four-terminal tile with terminals at angles $\theta_{tile} = \{0,\pi/2,\pi,3\pi/2 \}$): 

\begin{align}
\begin{split}
	\boldsymbol{z} \left( \theta_{tile} \right) &= \sum_{n=1}^{\infty} \frac{1}{n^2} \Big[ -\rho_{xx} \boldsymbol{S_n} -\rho_{xy} \boldsymbol{T_n} \cos(n \theta)  \\
	& - \rho_{xx} \boldsymbol{T_n} +\rho_{xy} \boldsymbol{S_n} \sin(n \theta) \Big] \nonumber 
\end{split} \\
	\qquad
	\rho_{xx} &=\frac{1}{en\mu} \quad \text{and} \quad \quad \rho_{xy}=\frac{B}{en} \nonumber \\
	\boldsymbol{U} &= \boldsymbol{\hat{Z}} \cdot \boldsymbol{I} + \boldsymbol{c} \quad \text{with} \quad \boldsymbol{\hat{Z}} =
	\begin{pmatrix}
	\boldsymbol{z} \left( 0 \right)^T \\
	\boldsymbol{z} \left( \pi/2 \right)^T \\
	\boldsymbol{z} \left( \pi \right)^T \\
	\boldsymbol{z} \left( 3\pi/2 \right)^T	
	\end{pmatrix}
\end{align}

The vectors $\boldsymbol{S_n}$ and $\boldsymbol{T_n}$ (given in SI) contain the Fourier coefficients of the current density entering the terminals. $\boldsymbol{c}=(c,c,c,c)^T$ is an undetermined constant that can be added to the electrostatic potential without changing the result. The resistor network (Fig: \ref{fig:2}\,a) is  formed by connecting the tiles with perfectly conducting wires. Accounting for current conservation and a continuous potential at the connections between tiles and boundary conditions (here chosen: 0\,V at the left bank and 1\,V at the right bank), a system of linear equations describing the network is derived and solved numerically using a sparse-matrix algorithm.

\begin{figure*}
%\twocolcaption
	\centering
  \includegraphics[width=178mm]{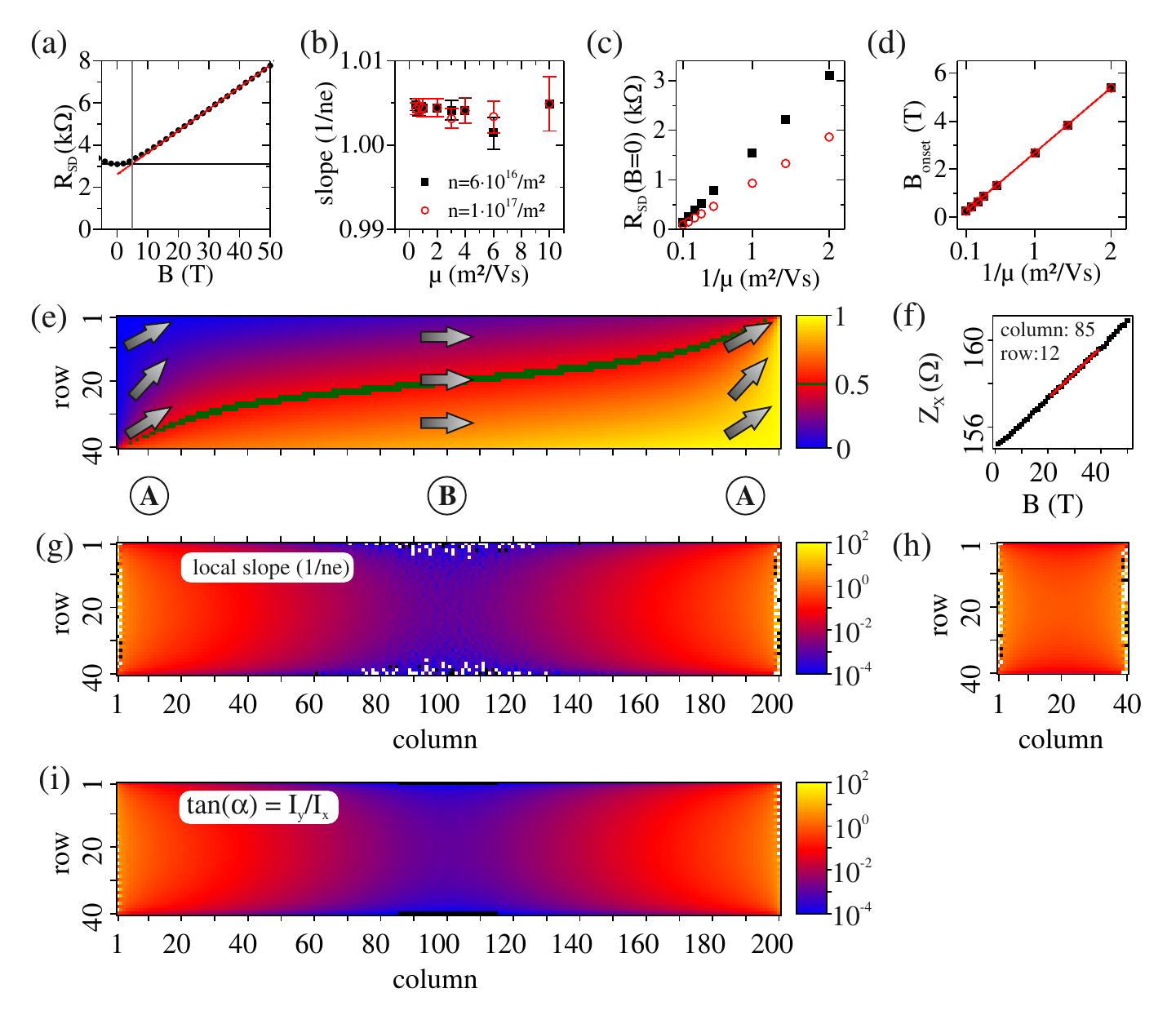}%
  \caption{(Color online): (a) Example for evaluation of characteristic MR parameters: slope at high fields, zero field resistance $R_{SD}(B=0)$ and onset field $B_{onset}$ for the 40$\times$200 homogeneous network. (b-d) Network parameters for homogeneous 40$\times$200  networks dependent on the mobility inside a tile $\mu$ calculated for two different charge carrier densities $n=6\cdot 10^{16}/\mathrm{m}^2$ (black squares) and $n=1\cdot 10^{17}/\mathrm{m}^2$ (red circles). (b) Slope of $R_{SD}(B)$, fitted from $30-50\, \mu B$, turns out to be nearly constant at a value of $1/ne$ independent of mobility. (c) Zero field resistance scales with $1/ \mu$ and also depends on charge carrier density via conductivity $\sigma = en \mu$. (d) Onset field is found proportional to $1 / \mu$ with $B_{onset} = (2.69 \pm 0.01)/\mu$, independent of charge carrier density $n$. (e) Potential map of a 40$\times$200 network at a magnetic field of 30\,T ($\mu = 1$). The arrows denote current direction (but not total current) at chosen tiles of the network. (f) Example for evaluation of the slope of mr locally calculated at a tile via $Z_{x}(B) = 2\cdot(\varphi_{right}-\varphi_{left})/(I_{left}+I_{right})$. (g) Logarithmic map of the local slope in units of ($1/ne$) for the network from (e). It divides in two regions: \textbf{A} near the electrodes and \textbf{B} in the center of the network, where the textbook Hall conditions are met. (h) Logarithmic map of the local slope for the corresponding square network, where only region \textbf{A} is present. (i) Logarithmic map of current ratio $I_x/I_y$ with $I_x$ averaged over left and right terminal and $I_y$ averaged over bottom and top terminal. \label{fig:3}} 
\end{figure*}

\section{Network Simulations: Results and Discussion}

\subsection{Homogeneous Networks}

In order to study the impact of our equivalent-circuit model on extended networks, we link up with the PL model that treats homogeneous, but finite networks, in which all discs are identical, arranged between two voltage-biased electrodes (source and drain). This model has lead to an understanding of linear MR, but barely of linear mr, as we will demonstrate.

The occurrence of finite CD fields $\alpha_\infty(\boldsymbol{r})$ (and the incomplete aspects of the PL model) can best be illustrated when choosing a matrix with a high aspect ratio (\# columns/\# rows). The MR calculated as a solution of the system of equations is displayed in Fig.\ \ref{fig:3} a: Beyond a certain threshold, the source-drain resistance $R_{SD}$ rises strictly linearly, without any saturation within our classical treatment. This simple curve can be characterized by an analysis of the slope, the minimal resistance $R_{SD} (B = 0)$ and the onset threshold $B_{onset}$ of linear behavior. The slope, expressed in units of $1/ne$ for two different values of $n$ clearly shows a behavior independent of $\mu$ (Fig.\ \ref{fig:3} b). $R_{SD} (B = 0)$ is trivially inversely proportional to the charge carrier mobility $\mu$ (Fig.\ \ref{fig:3} c). The threshold $B_{onset}$ is given by $\mu B \approx 1$, and independent of $n$ (Fig.\ \ref{fig:3} d). This set of parameters describes the phenomenology of linear MR.

Starting from $B = 0$ the potential map  evolves from a regular voltage drop along $x$ direction to a tilted landscape as in Fig. \ref{fig:3} e. For $B \gg 1/\mu$, this potential map has asymptotical behavior, i.e. changes only very slightly with increasing $B$. This can be rationalized by revisiting the concept of the Hall angle $\theta$, which is a measure of the ratio between the $y$ and $x$ components of the electric field $E$.  With the textbook results $\tan{\theta} = E_{y} / E_{x}= \mu B$ \cite{popovic}, the Hall angle saturates asymptotically beyond $\mu B \approx 1$, which freezes the potential landscape under voltage-biased conditions. The resulting potential map is rather insensitive to the tiling chosen for the finite-element analysis (see SI).

The potential map of the self-consistent solution reveals three spatial regions: the two areas close to the electrodes indicated with \textbf{A}  reproduce the diagonal equipotential line of Fig.\ \ref{fig:2} d. The central region \textbf{B} represents the situation in a real bulk material that is remote from electrode banks. Note that the potential map in this region is very similar to a standard textbook result of the Hall effect in strong magnetic fields with a Hall electric field that is approximately perpendicular to the current direction (Hall angle $\theta \approx 90^{\circ}$). The arrows in Fig.\ \ref{fig:3} e indicate the current direction in nine selected tiles. At such high fields ($\mu B \gg 1$) the current follows equipotential lines.

It is useful to introduce an analysis of local impedances $Z_x$ of each disc. We calculate $Z_x =U_x/I_x$ with $I_x = (I_{left}+I_{right})/2$ and $U_x$ being the potential drop along one disc. 
This scheme results in \textit{local} magneto-impedances that turn out to be linear in $B$, an example of which is shown in Fig.\ \ref{fig:3}\,f. An analysis of their slope indicates that it scales nicely with $1/ne$, with a proportionality constant that depends on the position. In region \textbf{A}, this proportionality is approximately unity, and the slope is $dZ_x (B)/dB \approx 1/ne$ (see Fig. \ref{fig:3} g). The spatial pattern of the local impedances resembles the PL case of a square matrix.

More importantly, in region \textbf{B} the slope is reduced by orders of magnitude. In other words, the parts remote from the electrode equipotential boundary condition have vanishing mr and, hence, do not contribute to linear MR. In order to corroborate this finding, it is instructive to elongate the matrix step by step, a procedure that inserts more and more contributions from region \textbf{B}. The outcome of this numerical experiment is unambiguous: the zero-field resistance trivially scales with the aspect ratio following Ohm's law in two dimensions, 
\begin {equation}
R_{SD} = R_\square \cdot \frac{l}{w}  \qquad (B=0),
\label{eqn:1}  
\end{equation}

with $l$ being the length and $w$ the width in appropriate units. In contrast, in the high-field limit, the linear-in-$B$ contribution is insensitive to geometry for rectangular homogeneous networks and its prefactor (slope) is simply $1/ne$ (Fig. \ref{fig:3}\,c). It is therefore an electrode-induced MR rather than a specific mr. The overall MR in the high-field limit reads

\begin {equation}
R_{SD} \approx \underbrace{R_\square \cdot \left( \frac{l}{w}-1 \right)}_{\text{\textbf{B}}} + \underbrace{\frac{B}{ne}}_{\text{\textbf{A}}} \qquad (\mu B\gg 1).
\label{eqn:2}  
\end{equation}

Note that the two terms stand for region \textbf{A} and \textbf{B}, although these regions are not sharply separated in space. The PL model treats only the latter term, while linear mr in experiments was observed in four-terminal measurements, i.e. in geometries that are only sensitive to the first term \cite{Kisslinger}.

In this section we have so far argued in the conceptual framework of the finite-element/PL model. At this stage, however, we can easily link it with the individual tile model described in section \ref{sec:generating_mechanism}. In particular, the resistivity of the individual tile, $R_x^\prime = U_x/I_x \propto \tan(\alpha_\infty) \frac{B}{ne}$ at high magnetic fields (cf. formula \ref{eqn:4}), with $\alpha_\infty$ determined by the environment, can now immediately be linked to the local impedances $Z_x(\boldsymbol{r})$ calculated with the finite-element model.

 One can simply calculate the CD field $\alpha_\infty(\boldsymbol{r})$ within the finite-element model for each element, and map it as $\tan[\alpha_\infty(\boldsymbol{r})]=I_y/I_x$  for the homogeneous network (Fig. \ref{fig:3} i). The coincidence of the maps in Fig. \ref{fig:3} g and i (with significant differences only at the edges where translational symmetry is strongly violated) gives convincing evidence that the equivalent circuit model,  in conjunction with the CD field $\alpha_\infty(\boldsymbol{r})$ that has to be derived from geometry, is the relevant point of view to understand linear magnetoresistivity.

\subsection{Inhomogeneous Networks}

\begin{figure*}
%\twocolcaption
	\centering
  \includegraphics[width=0.85\textwidth]{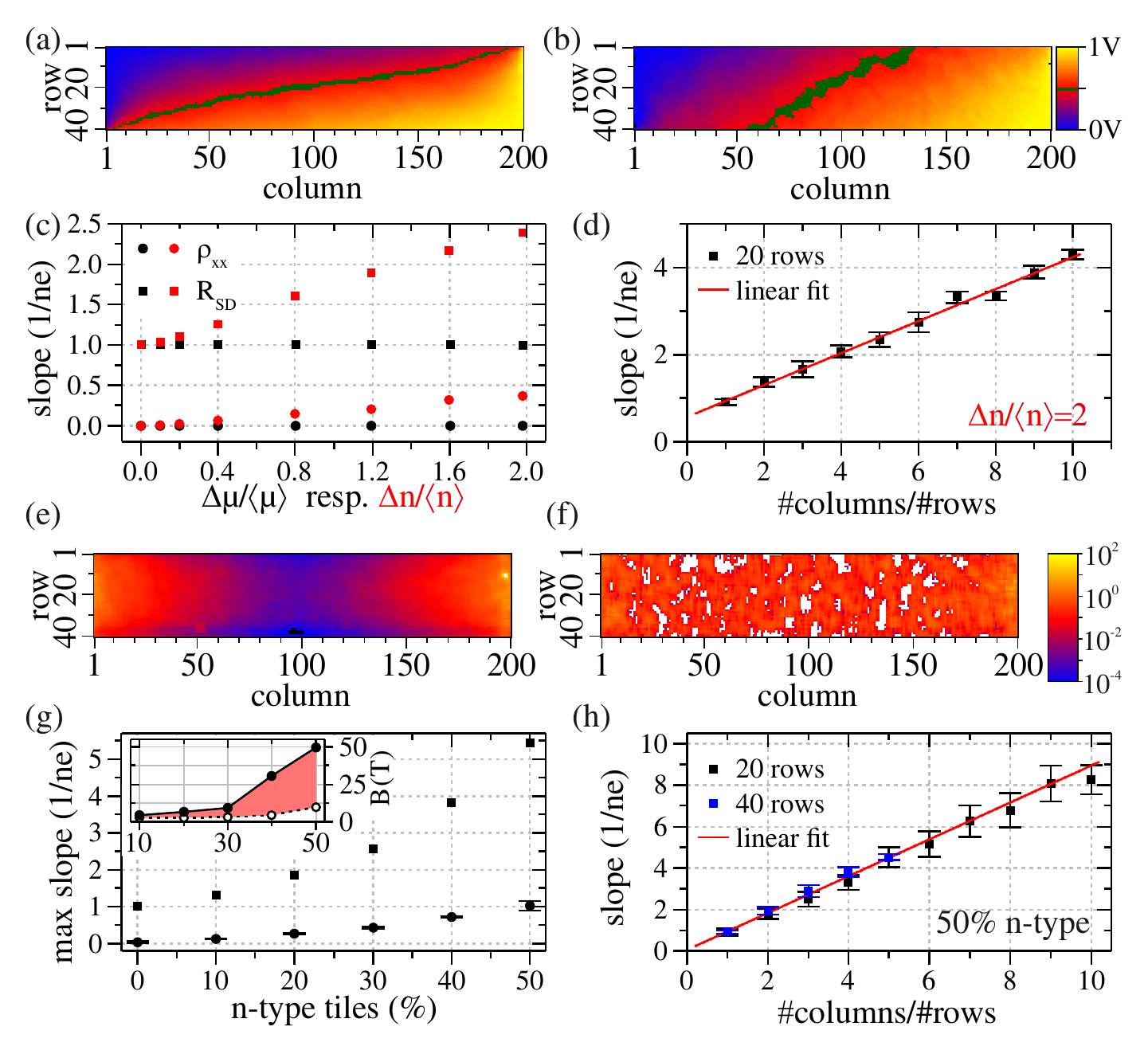}%
  \caption{(Color online): (a and b) Potential maps for disordered 40$\times$200 networks at a magnetic field of $30\,$T. A flat distribution with mean value $\left\langle x\right\rangle$ and total width $2 \cdot \left\langle x\right\rangle$ was used for these maps. (a) shows the influence of disorder in $\mu$ ($n$ fixed). In (b) the charge carrier density $n$ was varied ($\mu$ fixed). (c)  Slope of linear MR/mr (in units of $1/ne$ with $n=6\cdot10^{16}/\mathrm{m}^2$) as a function of the total width of disorder distribution for $\mu$ or $n$ disorder. The squares indicate the slope of the source-drain-resistance $R_{SD}$, whereas the circles show the slope of the bulk resistivity $\rho_{xx}$. (d) Slope of the network resistance (see (c)) under variation of aspect ratio of the network. The values for the network with 20 rows were averaged over five configurations for each aspect ratio. A linear fit yields a slope of $(0.37 \pm 0.01)$ and an offset of $(0.57 \pm 0.05)$. (e and f) Map of CD field. The plotted quantity is $\tan[\alpha(\boldsymbol{r},B=30\,\mathrm{T})-\alpha(\boldsymbol{r},B=0)]$, i.e.the field-induced change of the local current direction with respect to the local electric field. The values are averaged over a square of $5 \times 5$ tiles, negative values are displayed in white. Panel (e) corresponds to the case of $\mu$ disorder (see panel (a)) and panel (f) to the case of $n$ disorder (see panel (b)). (g) Maximum slope of the source-drain-resistance (black squares) and the resistivity (black circles), in region \textbf{B}, of 40$\times$200 networks in dependence of the fraction of $n$-type discs (for evaluation details see SI). For the simulation, charge carrier density $n_{n,p}=6\cdot10^{16}/\mathrm{m}^2$ and mobility $\mu = 1 \mathrm{m}^2/\mathrm{Vs}$ were fixed and just the sign of the Hall-coefficient varied statistically between the different tiles. Inset: Lower (open circles) and upper (filled circles) limit of the magnetic field region in which the resistivity is found to be linear-in-$B$. (h) Slope of $R_{SD}$ for different aspect ratios of networks with equal fraction of $n$- and $p$-type tiles. Each point represents an average over five configurations. A linear fit yields a slope of $(0.89 \pm 0.02)$ and an offset of $(0.05 \pm 0.05)$. \label{fig:4}} 
\end{figure*}

We have shown that in homogeneous conductors linear mr is built in, however with zero amplitude when being remote from electrodes (region {\bf B}). Linear MR and mr, however, have been suspected as originating from disorder. Conceptually, the generating mechanism presented in chapter \ref{sec:generating_mechanism} relies on translational invariance, i.e. can not be applied in the inhomogeneous case. This is why we consider inhomogeneities in the framework of the the finite-element model \cite{PL1,PL2}. It provides a tool  to continuously enhance the degree of disorder and to extend the underlying physics to the disordered regime. As we will see, disorder goes along with a non-vanishing $\tan [\alpha_\infty(\boldsymbol{r})]$ field.

Inhomogeneities are introduced \textit{via} a stochastic choice of parameters for the individual tiles. We have chosen disorder of the charge carrier mobility or, alternatively, charge carrier density, expressed by an equal distribution of total width $\Delta \mu$ or $\Delta n$ centered around an average value $\left\langle \mu \right\rangle$ or $\left\langle n \right\rangle$, respectively. This procedure allows to independently control the continuous evolution from weak to strong inhomogeneities.  Fig. \ref{fig:4} a and b show the effect on the potential map at high magnetic field ($B = 30\,$T), and under a stochastic variation of $\mu$ and $n$, respectively. Obviously, the $U = 0.5\,$V equipotential line becomes distorted as compared to Fig. \ref{fig:3}e. Note that $n$ disorder tilts the equipotential line much stronger than $\mu$ disorder. This is not surprising, as the local Hall resistance scales with $1/ne$. An analysis of the source-drain resistance $R_{SD}$ of the device as a function of disorder $\Delta\mu$ and $\Delta n$ is shown in Fig. \ref{fig:4} c. It turns out that the effect of charge carrier density variations $\Delta n$ is immediately acting on linear MR, whereas $\Delta \mu$ is barely influential (cf. \cite{magier}). The effect of the boundaries can be eliminated by investigating linear mr of $\rho_{xx}$ obtained in the bulk region \textbf{B}. Here, the linear-in-$B$ contributions start from zero, and show a rapid increase when $n$ disorder is applied. This can again be traced back to the CD fields. Fig. \ref{fig:4} e and f show the CD map, more explicitly $\tan [\alpha_\infty(\boldsymbol{r})]$ for the case of $\mu$ and $n$ disorder, respectively. The difference is stunning: $\mu$ disorder does not cant the current with respect to the bias direction in the middle of the simulated device, whereas $n$ disorder efficiently creates finite CD fields, which are distributed over the whole area and let even the near-electrode regions (region \textbf{A}) unstructured in comparison to Fig. \ref{fig:4} e. Consequently, charge carrier density disorder delivers a contribution to the bulk specific resistivity, i.e. it creates rather a linear magneto\textit{resistivity} (mr) than a linear magneto\textit{resistance} (MR): every square of the two-dimensional sample contributes linear mr, a fact that is corroborated by a variation of the aspect ratio ($AR$) of the network (see Fig. \ref{fig:4} d): The longer the sample, the larger the linear MR of the sample. The offset stems from region \textbf{A}, i.e. from the electrodes. The latter effect next to the electrodes can be fully traced  back to the charge density contrast between the tiles (finite $n$) and the equipotential electrodes ($n = \infty$).

Note that in Ref. \cite{PL1}  disorder was treated in a square device geometry, i.e. fully dominated by region \textbf{A}. The authors made the
less appropriate choice of considering $\mu$ disorder, for which they had to assume enormously high values of $\left\langle \mu \right\rangle/\mu \gg 1$ to overcome the strong influence of boundary conditions. In this limit they included negative values, i.e. a mixture of electrons and holes with statistical emphasis on vanishing mobilities. This is an extreme,  barely realistic case the essence of which we treat now in a simpler bipartite ensemble: instead of a gaussian distribution around {$\mu\approx 0\,$} \cite{PL1},
 we assume two sorts of tiles with well-controlled homogeneous charge carrier mobilities and -density, but with
positive and negative charge carriers (p-type and n-type tiles),  randomly distributed in a rectangular four-terminal network. Such bipartite materials have been treated in model calculations \cite{stroud1,Guttal2005} and have revealed linear MR at exact electron/hole balance. In our finite-element treatment, starting off with a percentage of n-type tiles from zero, i.e. from a homogeneous network, the linear slope increases rapidly when more and more n-type tiles are introduced (see Fig. \ref{fig:4} g). In particular, the slope of the mr $\rho_{xx}$ increases and reaches values as big as $1/ne$ per square, indicating a very efficient generation of linear mr, also illustrated by a variation of aspect ratio of the slope of linear mr (Fig. \ref{fig:4} h).

The slope of the resistance consequently reaches a maximum as high as $5/ne$ when the number of n- and p-type tiles is equal, which reflects the aspect ratio of the simulated device (40$\times$200 tiles). For mixtures in between 0 and 50\%, the mr saturates at finite $B$, recalling the traditional case of a saturating mr in the presence of electrons and holes in a homogeneous medium \cite{popovic} and linear mr only appears in a limited field region \cite{Guttal2005} (see inset in Fig. \ref{fig:4} g). For exactly 50\%, the case resembles to \cite{gornyi}. Note that also in the disordered case, the potential landscape freezes with increasing $B$ field for rectangular device geometries.

This elucidates the generating mechanism for linear mr in disordered materials: the disorder provides a stochastic mixture of longitudinal and Hall electric fields. Consequently, the formation of a macroscopic Hall field (in which the Lorentz force is close to zero) is impossible, but instead locally varying mismatch between the current and the bias direction is enforced. This can be understood as finite, stochastically varying values of $\alpha_\infty$ at each tile. The most important point, however, is the interplay of the \textbf{local} resistivity matrix with the \textbf{nonlocal} potential map.
%is the effective medium that shares the same linear $B$-dependence of the resistivity as the the current-voltage relation of the tile itself. 
Our analysis of the inhomogeneous network resistance (Fig. 4) clearly shows that the feedback mechanism by the effective medium expressed in eq. \ref{eqn:4} overcompensates the loss of translational invariance when disorder is turned on and stabilizes linear magnetoresistivity both locally and globally. In that sense, linear mr is rather the rule than the exception when an additional nonlocal mechanism provides locally varying current directions in strong $B$ fields, for example next to a metallic electrode or in a disordered medium.

\section{Critical Review}\label{sec:CriticalReview}

In the light of our findings the slope of linear magnetoresistance and linear magnetoresistivity is inversely proportional to $n$, but insensitive to $\mu$ in the high field limit $\mu B > 1$. This simple finding has been hidden both in theoretical and in experimental work by the commonly used plot of $\rho(B)/\rho(0)$ that explicitly removes the $n$-dependence. Note that within simple Drude formulae the division by $\rho (0)$ is equivalent to multiplication with $\mu \cdot n \cdot e$, which brings $\mu$ artificially into play and was subsequently related to the slope of linear MR \cite{PL1,PL2}. When further including mobility fluctuations, electron-hole mixing had to be assumed (expressed as $\Delta \mu \gg \left\langle \mu\right\rangle$) in order to see an effect on linear MR, which is again rather charge disorder than mobility disorder.

As a consequence of this treatment, the slope of linear mr has been assigned to mobility fluctuations, which was misleading and has strongly influenced the interpretation of experiments \cite{WangSciRep,WangAPL2014,Wang2014,Hu2008,Johnson2010,Narayanan,Yan2013,Delmo,Aamir}. The generating mechanism is much more sensitive to charge carrier density fluctuations \cite{Porter}, as can be seen in Fig. \ref{fig:4} c. 

We know from experiment that additional mechanisms exist that generate finite CD fields, and, consequently, linear mr. In the case of  bilayer graphene, charge carrier density is fixed by the epitaxially defined surface \cite{Kisslinger}, and the mosaic tiling defined by 50 nm distant partial dislocations is so small that the charge carrier mobility inside a tile is ill-defined (tile size and mean free path coincide). Nevertheless, we robustly find a strong linear mr in each and every of about 30 samples investigated. This suggests that structural disorder prepatterns the current map and therefore creates finite  $\tan [\alpha_\infty(\boldsymbol{r})]$ CD fields. The distorted current/potential map, together with the strong and robust effective medium argument, is sufficient to generate linear mr. This is presumably why topological materials tend to display linear mr as well \cite{Narayanan,Roychowdhury,WangSciRep}.

The importance of disorder and finite CD fields, in coincidence with high overall mobilities, provide the link to granular materials, in which linear mr has first be observed. Note that magnetoresistance is essentially a 2D phenomenon. When transferring the model to a 3D material, the generating mechanism, however, is expected to remain valid.

 \section{Summary}

Key to the understanding of linear magnetoresistivity of a conductor in classically strong magnetic fields is an effective medium argument, which can be traced back for \textit{homogeneous materials} to the simple equivalent circuit in Fig. 1 g: When the feedback of Hall currents is via the same material, only $R\propto B$ is a self-consistent solution at high magnetic fields. However, an additional \textit{nonlocal} mechanism is required, which cants the current with respect to the applied biasing electric field by an angle  $\alpha_\infty(\boldsymbol{r})$ and thus create a finite current distortion field. This, in apparent contradiction, can only be provided in \textit{inhomogeneous materials}. In order to understand such a non-local canting mechanism, a finite-element analysis is well suited. One outcome is that charge carrier density fluctuations are most efficiently creating finite current distortion fields, whereas mobility fluctuations are only weakly influential. A special case of charge density contrast occurs when an extended metallic electrode is added, which is particularly influential for linear mr. The simulations further prove that the generating mechanism is so robust that its effect survives even when homogeneity is lifted.  

Linear mr can be best observed when (i) charge density is low, leading to large local Hall effect, (ii) The overall mobility is high, such that the linear MR can be observed at small magnetic fields ($\mu B$ exceeds unity), and (iii) a nonlocal mechanism that creates a significant current distortion field  $\alpha_\infty (\boldsymbol{r})$. For the "simple" conductor considered here, the effect is generic. 
The effect is, however, obscured or replaced in other parameter regimes, including the low-temperature Landau Quantization regime, perfect electron-hole balance, or complex Fermi surfaces.

Altogether, linear mr is rather the rule than the exception when inhomogeneities distort current pathways in a simple classical low-density conductor. 

\vspace*{1em}

\begin{acknowledgments}
The work was carried out in the framework of the SFB 953 and SPP 1243 of the Deutsche Forschungsgemeinschaft (DFG). 
\end{acknowledgments}

\bibliography{bibliography-origin}

\clearpage

\setcounter{figure}{0}
\setcounter{section}{0}
\large{\textbf{SUPPORTING INFORMATION}
\normalsize

\section{Derivation of impedance matrix}\label{sec:impedance_matrix}
The electrostatic potential $\varphi$ at the edge of a circular disc with radius $R$ results from the Laplace equation $\Delta \varphi = 0$ and is given by: 
\begin{align}
	\varphi (\theta_{tile},R) &= c + \cdot \sum_{n=1}^{\infty}{R^n \left[ A_n \cos(n\theta_{tile}) + B_n \sin(n\theta_{tile}) \right]}
\end{align}
The calculation is performed in polar coordinates with $\varphi$ written as a Fourier series. Writing the current density $\boldsymbol{j}=j_r \cdot \boldsymbol{e_r}$ (there is only a radial component assumed), which is inserted into the tile also as a Fourier series one obtains in the case of four contact terminals at the tile: 
\begin{align}
\begin{split}
	j_r (\theta_{tile}&,R) = \sum_{k=1}^{4} \Bigg\{ \frac{I_k }{2 \pi R} \\
	&+ \frac{I_k}{\pi R} \sum_{n=1}^{\infty}{ \frac{1}{n} \left[ S_{n.k} \cos(n\theta_{tile}) + T_{n,k} \sin(n\theta_{tile}) \right] } \Bigg\} 
	\end{split}\\
	\begin{split}
	\quad S_{n,k} &= \frac{1}{\delta_{t,k}} \Bigg\{ \sin\left[ n \left( \theta_k + \frac{\delta_{t,k}}{2} \right) \right] \\
	&\hspace{0.3\columnwidth} - \sin\left[ n \left( \theta_k - \frac{\delta_{t,k}}{2} \right) \right]  \Bigg\}
	\end{split}\\
	\begin{split}
	T_{n,k} &= \frac{1}{\delta_{t,k}} \Bigg\{ \cos\left[ n \left( \theta_k - \frac{\delta_{t,k}}{2} \right) \right] \\
	&\hspace{0.3\columnwidth} - \cos\left[ n \left( \theta_k + \frac{\delta_{t,k}}{2} \right) \right] \Bigg\}
	\end{split}
\end{align}
The index $k$ denotes the terminal. The terminals are located at the angle $\theta_k$, with an opening angle $\delta_{t,k}$ and a current $I_k$ entering the terminal (see Fig. 1\,b). The current density is assumed to be constant over the whole opening angle of each terminal. Comparing coefficients of the left and right hand side of Ohms law $- \boldsymbol{\hat{\sigma}} \cdot \boldsymbol{\nabla} \varphi  = \boldsymbol{j}$, the impedance matrix $\boldsymbol{\hat{Z}}$ given in the main manuscript can be calculated:
\begin{align}
	\begin{split}
	\boldsymbol{z} \left( \theta_{tile} \right) &= \sum_{n=1}^{\infty} \frac{1}{n^2} \Big[ -\rho_{xx} \boldsymbol{S_n} -\rho_{xy} \boldsymbol{T_n} \cos(n \theta)  \\
	& - \rho_{xx} \boldsymbol{T_n} +\rho_{xy} \boldsymbol{S_n} \sin(n \theta) \Big] \nonumber 
\end{split} \\
	\qquad
	\rho_{xx} &=\frac{1}{en\mu} \quad \text{and} \quad \quad \rho_{xy}=\frac{B}{en} \nonumber \\
	\boldsymbol{U} &= \boldsymbol{\hat{Z}} \cdot \boldsymbol{I} + \boldsymbol{c} \quad \text{with} \quad \boldsymbol{\hat{Z}} =
	\begin{pmatrix}
	\boldsymbol{z} \left( 0 \right)^T \\
	\boldsymbol{z} \left( \pi/2 \right)^T \\
	\boldsymbol{z} \left( \pi \right)^T \\
	\boldsymbol{z} \left( 3\pi/2 \right)^T	
	\end{pmatrix}\\
	\text{with}\\
	\boldsymbol{S_n} &= \frac{1}{\pi}
	\begin{pmatrix}
		S_{n,1} \\
		S_{n,2} \\
		S_{n,3} \\
		S_{n,4}
	\end{pmatrix} \\
	\boldsymbol{T_n} &= \frac{1}{\pi}
	\begin{pmatrix}
		T_{n,1} \\
		T_{n,2} \\
		T_{n,3} \\
		T_{n,4}
	\end{pmatrix}
	\quad \text{,}
\end{align}
where $\boldsymbol{c} = (c,c,c,c)^T$ is a constant added to all components of the potential vector $\boldsymbol{U}$.

\section{Resistance of the feedback circuit}
The resistance of the feedback circuit (see Fig. 1) is calculated using simple formulae for the Hall effect in the tile. The resistance tensor
\begin{align}
	\begin{pmatrix}
		U_x \\
		U_y
	\end{pmatrix} =
	\begin{pmatrix}
		\rho_0 & B/ne \\
		-B/ne & \rho_0
	\end{pmatrix} \cdot
	\begin{pmatrix}
		I_x \\
		I_y
	\end{pmatrix}
\end{align}
leads to the following equations:
 %The Hall-voltage $U_y$ is calculated via the longitudinal current $I_x$ and is equal to the voltage drop at the feedback resistor $R_{eff}$ via Kirchhoff`s loop rule. The longitudinal voltage drop $U_x = U$ is given by the longitudinal current $I_x \cdot R_\square$ and the increase of voltage via the Hall-current $I_y\cdot B/ne$ (under voltage biased conditions this leads to a reduction of current). Altogether this reads:
\begin{align}
	U_x &= \rho_0 \cdot I_x + \frac{B}{ne} \cdot I_y = U \\
	U_y &= -\frac{B}{ne} \cdot I_x + \rho_0 \cdot I_y = - R_{eff} \cdot I_y \\
\end{align}
Here it was used that the voltage drop at the feedback resistor $R_{eff} \cdot I_y$ equals the voltage drop $U_y$ at the tile (with inverted sign) \textit{via} Kirchhoff`s loop rule. This results in a resistance $R_x^\prime = U_x/I_x$
\begin{align}
	R_x^\prime &= \rho_0 + \frac{(B/ne)^2}{\rho_0 + R_{eff}} \quad \text{.}
\end{align}
Self-consistency is achieved by choosing a linear-in-$B$ feedback resistor $R_{eff}$, which accounts for the environment of the tile in the conductor. This resistance is given by the resistance of the tile itself and by the environment. The circuit resistance $R_x^\prime$ is hence given by:
\begin{align}
	R_{eff} &= R_{eff,0}+ \frac{1}{\tan(\alpha_\infty)} \cdot \frac{B}{ne} \\ 
	R_x^\prime &= \rho_0 + \frac{(B/ne)^2}{\rho_0 + \left[ R_{eff,0}+ \frac{1}{\tan(\alpha_\infty)} \cdot \frac{B}{ne} \right]}
\end{align}
In the high field limit $B \gg \tan(\alpha_\infty) R_{eff,0} n e \approx \mu B$ the resistance reduces to:
\begin{align}
	R_x^\prime &=   \tan(\alpha_\infty) \cdot \frac{B}{ne} 
\end{align}
Where $\tan(\alpha_\infty)$ adjusts the strength of the linear contribution to $R_x^\prime$ and can be identified as the current ratio $I_y/I_x$ in this limit. A fit of this model to the current ratio, which was calculated for the 40$\times$200 network (see main manuscript), is shown for completeness in Fig. SI \ref{fig:SI1}. The formula for the current ratio in this model is given by:
\begin{align}
	\frac{I_y}{I_x} = \frac{B/ne}{R_{eff,0}+ \frac{1}{\tan(\alpha_\infty)} \cdot \frac{B}{ne}} \label{eqn:current_ratio}
\end{align}
Where the $\rho_0$ was absorbed to $R_{eff,0}$ for the fitting procedure.

\begin{figure*}[t!]
    \centering
        \includegraphics[scale=1.0]{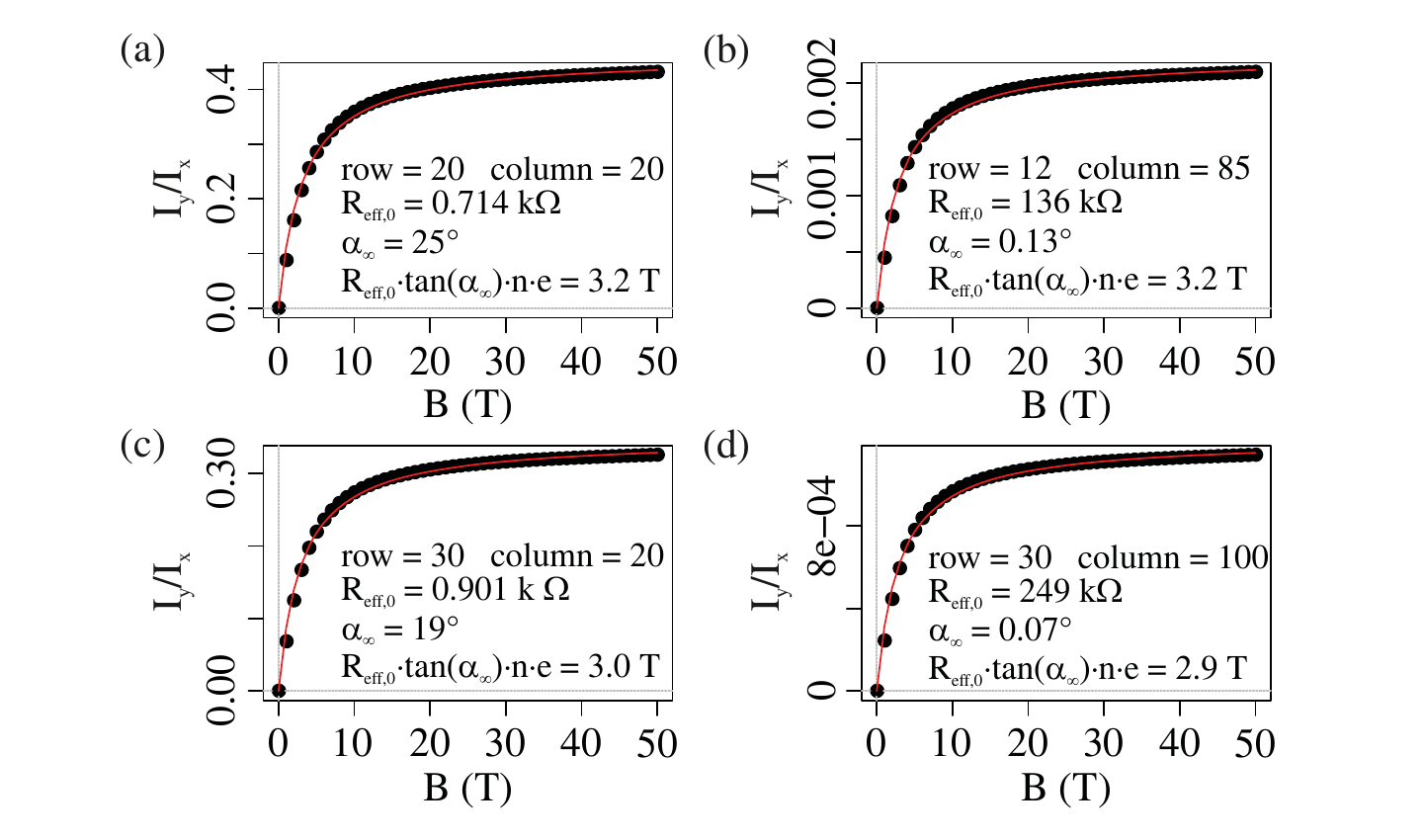}
                        \caption{\textbf{SI} (Color online) Panel (a-d): Current ratio $I_{top}/I_{left}$ (black circles) plotted for four selected tiles. The red line is a fit of formula \ref{eqn:current_ratio} to the model, which yields the parameters $\alpha_\infty$ and $R_{eff,0}$. These parameters depend on the position of the tile. Whereas the field scale $\tan(\alpha_\infty) R_{eff,0} n e$ is nearly independent of position. We attribute the remaining difference between the fit and the simulated data to a systematic error due to the finite size of the network.}
    \label{fig:SI1}
\end{figure*}

\section{Further network geometries}
In order to study the sensitivity of the linear mr to the model chosen, in which each rectangular Hall tile has four terminals to its neighbors, we carried out simulations in further configurations: (i) four-terminal tiles which are tilted by 27\textdegree and 45\textdegree with respect to the bias direction (Fig. SI \ref{fig:SI2} e) (ii) three-terminal tiles which form a honeycomb lattice (Fig. SI \ref{fig:SI2} c). Fig. SI \ref{fig:SI2} a and b show potential maps thus obtained, all for homogeneous networks (all tiles are equal). It turns out that current maps (not shown) depend strongly on the geometry chosen. However, the resulting potential maps are barely distinguishable (see Fig. SI \ref{fig:SI2} a,b), as well as the slope of linear MR (Fig. SI \ref{fig:SI2} d). We conclude that, at least for homogeneous networks of sufficient size, the finite-size analysis of the potential map as well as the linear MR are insensitive to the tiling chosen. The three terminal geometry (Fig. SI \ref{fig:SI2} c) was build up of tiles with three terminals at angles $0^{\circ}$, $120^{\circ}$ and $240^{\circ}$. The calculation of the impedance matrix follows exactly the scheme presented above in section \ref{sec:impedance_matrix}.\\
The tilted geometry (Fig. SI \ref{fig:SI2} e) was implemented by cutting it out of a larger network. The insulating edges of the tilted network were formed by high ohmic tiles ($\approx 10\,\mathrm{M} \Omega$), the metallic electrodes were realized by low ohmic tiles ($\approx 60\,\mathrm{m} \Omega$) with high charge carrier density.

\begin{figure*}
  \includegraphics[width=0.7\textwidth]{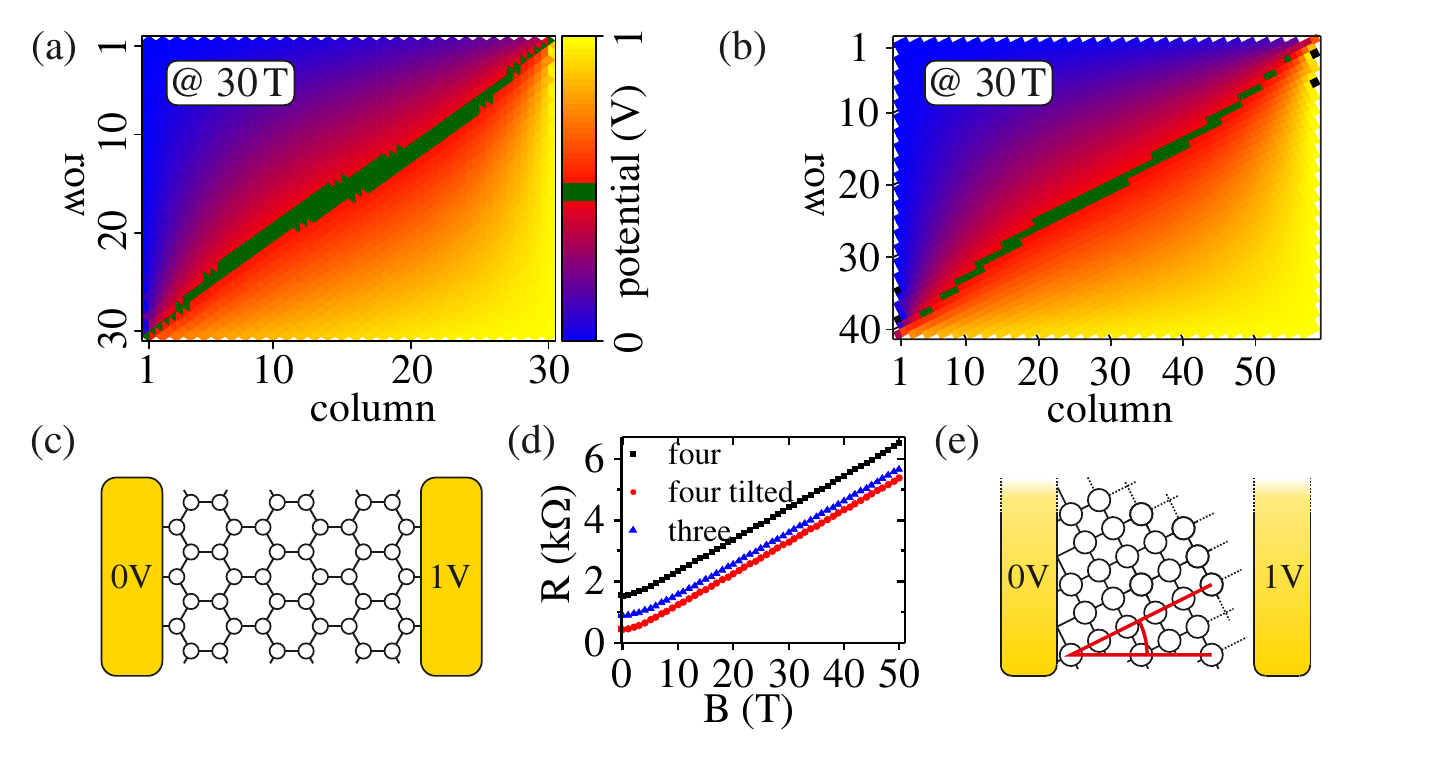}%
  \caption{\textbf{SI} (Color online): (a) Potential map for a homogeneous network with hexagonal geometry (30$\times$30 hexagons). A sketch of the network geometry (3$\times$3 hexagons) can be found in (c). (b) Potential map of a tilted (27\textdegree) four-terminal-network (for details see SI). A sketch of the network geometry is shown in (e). (d) Network resistance for the different geometries. The zero field resistance depends on geometry as well as on aspect ratio. The slope is found to be independent of geometry at a value of $\approx 1/ne$. \label{fig:SI2}} 
\end{figure*}

\section{Inhomogeneous networks}
The evaluation scheme for the disordered networks, presented in Fig. 4 c, is shown in Fig. SI \ref{fig:SI3}, but is representative for whole range from $\Delta x = [0;2\left\langle x\right\rangle]$ with $x=n,\mu$. The data presented is for the case $\Delta n / \left\langle n \right\rangle = 2$ and $\Delta \mu / \left\langle \mu \right\rangle = 2$ respectively. The same network configuration as for Fig. 4 a and b was used for this example evaluation. The slope in the specific resistivity $\rho_{xx}$, presented in Fig. 4 c, was calculated by averaging the slopes, which turned out of the evaluation scheme Fig. SI \ref{fig:SI3} (a-d). Therefore a window of width 60 columns was shifted along the bottom row of the network in region \textbf{B}. The specific resistance $\rho_{xx}$ was calculated via
\begin{align}
	\rho_{xx} = \frac{\Delta U}{I} \cdot \frac{\#rows}{\#columns} \qquad \text{.}
\end{align}
The region investigated started at column 55 and ended at column 145 (40$\times$200 network), this results in 31 values for the slope which were averaged. For the case of increasing $n$ disorder the onset field $B_{onset}$ is shown in Fig. SI \ref{fig:SI3} e. This onset field was calculated as the intersection point of the zero field resistivity $\rho_{xx}(B=0)$ with a linear function fitted to the the high field magnetoresistivity. The onset field decreases trivially with the inceasing $\Delta n/n$, as the slope is also inceasing in this case. The calculated source-drain-resistances $R_{SD}$ for the variation of aspect ratio (corresponding to Fig. 4 d) are displayed in Fig. SI \ref{fig:SI6} a.

\subsection{Networks with mixing of electrons and holes}
The networks presented in Fig. 4 e and f were calculated using tiles with either electron transport ($n$-type) or hole transport ($p$-type). These tiles were mixed together with a random position in the network. The tiles had all the same mobility of $\mu_{p,n} = 1\,\mathrm{m}^2/\mathrm{Vs}$ and the same charge carrier density denoted as $n_{p,n} = 6\cdot10^{16}/\mathrm{m}^2$, but they had different sign of the Hall-coefficient.\\
In the case of Fig. 4 e the fraction of tiles with $n$-type transport was varied in 40$\times$200 networks. The evaluation scheme for the derivation of the slope of $\rho_{xx}(B)$ and the width of the linear region in the magnetic field $B$ is presented in Fig. SI \ref{fig:SI4} for an example window (column 75 to 135) and for the different percentage of $n$-type tiles. It is done the following way:
\begin{enumerate}%[1.]
	\item Calculate the symmetric part of $\rho_{xx}$.
	\item Differentiate $\rho_{xx}$ numerically.
	\item Slope is taken here as the maximum of the derivative.
	\item The lower and upper limits of the linear region is set at the magnetic field value, where the derivative has dropped about $20\,$\% of the maximum value.
\end{enumerate}
To make the evaluation for the different percentages of $n$-type tiles better comparable the data for the evaluation Fig. SI \ref{fig:SI4} is plotted together in Fig. SI \ref{fig:SI5} for all percentages of $n$-type tiles. The parameters for the source-drain-resistance $R_{SD}$ were calculated the same way, note that in this case there is always an offset of $1/ne$ in the derivative of $R_{SD}$ caused by region \textit{A}.\\
e evaluation Fig. SI \ref{fig:SI4} is plotted together in Fig. SI \ref{fig:SI5} for all percentages of $n$-type tiles. The parameters for the source-drain-resistance $R_{SD}$ were calculated the same way, note that in this case there is always an offset of $1/ne$ in the derivative of $R_{SD}$ caused by region \textit{A}.\\
The variation of the aspect ratio Fig. 4 f was done using equal fractions of $n$- and $p$-type tiles. The slope of the resulting $R_{SD}$ was calculated by a linear fit to each curve in the range from 20$\,$T to 40$\,$T. The values for the same aspect ratio of the network and for the same number of rows were averaged afterwards, the error of these averaged values (given in the main manuscript) is just the standard deviation of these values. The calculated source-drain-resistances $R_{SD}$ for the variation of aspect ratio are displayed in Fig. SI \ref{fig:SI6} b and c.

\begin{figure*}[t!]
    \centering
        \includegraphics[scale=1.0]{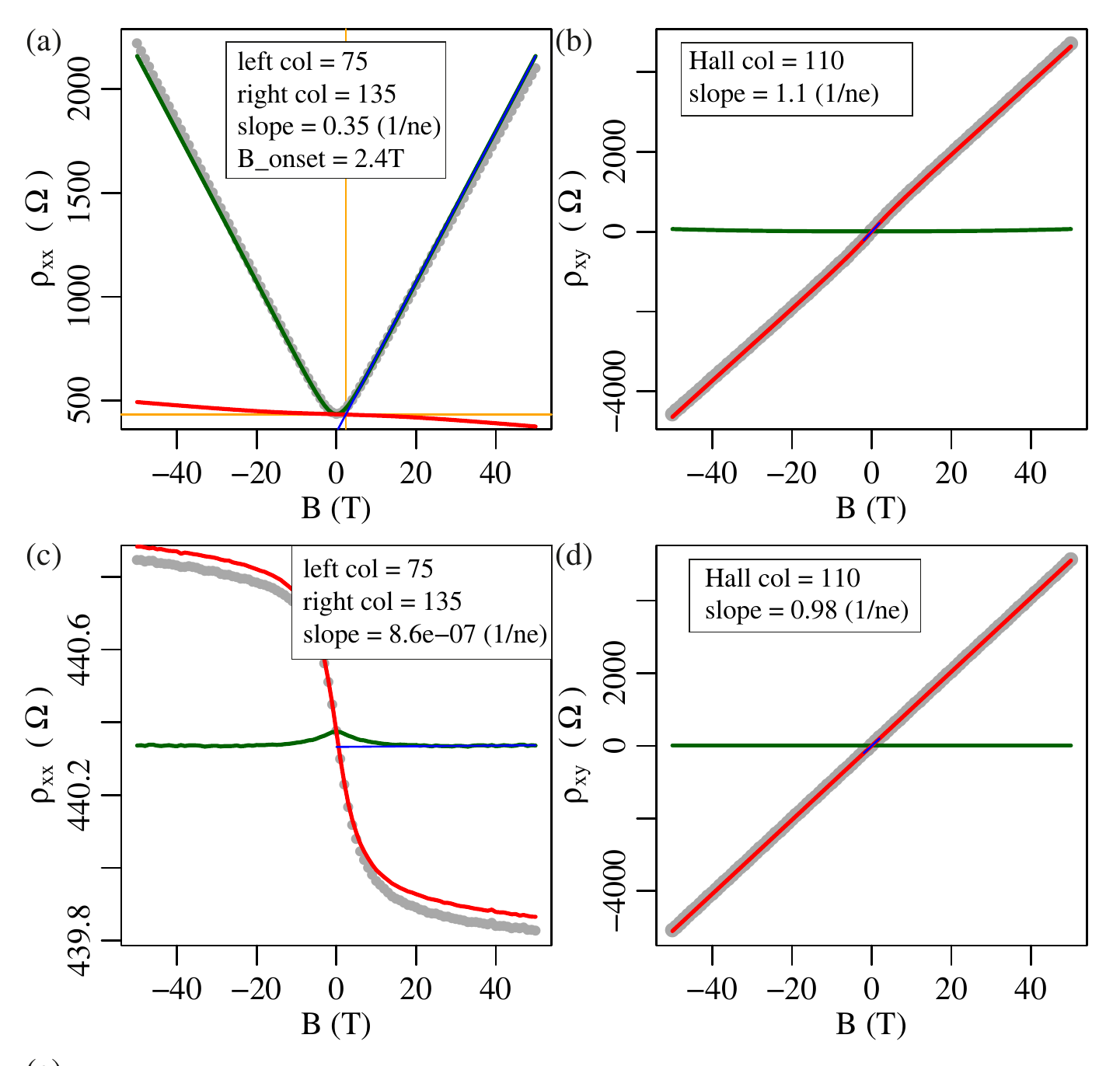}
                        \caption{\textbf{SI} (Color online) Panel (a) and (b) correspond to $n$ disorder, Panel (c) and (d) to $\mu$ disorder. In both cases the total width of the distribution centered around the mean value $\left\langle x \right\rangle$ is twice the mean value, wich is the maximum width if one does not allow for mixing of electrons and holes. (a) Four-point resistivity $\rho_{xx}$ (grey circles). Voltage difference was calculated at columns 75 and 135 as four-point-probes at the bottom of the network. Green line ist the part of $\rho_{xx}$ symmetric in $B$ and red line is the antisymmetric part in $B$ (with the zero fiel resistivity as offset). A linear fit (blue line) to the symmetric part for fields $B > 30 \,$T yields a slope of $0.35/ne$ with $n=6\cdot10^{16}/\mathrm{cm^2}$. Vertical yellow line shows the evaluation of the onset field $B_{onset}$ as crossing point between linear fit at high fields and zero field resistance. (b) Hall-resistivity $\rho_{xy}$ evaluated at column 110 (grey circles). A linear fit to the antisymmetric part (red line, green line shows the symmetric-in-$B$ part) at low fields yields a Hall constant $1.1/ne$ with $n=6\cdot10^{16}/\mathrm{cm^2}$. Panel (c) and (d) show the same evalutaion of parameters as panel (a) and (b) respectively, but for the case of mobility disorder. (e) Onset field $B_{onset}$ of $\rho_{xx}$ for the case of increasing $n$ disorder.}
    \label{fig:SI3}
\end{figure*}

\begin{figure*}
				\centering
        \includegraphics[scale=0.8]{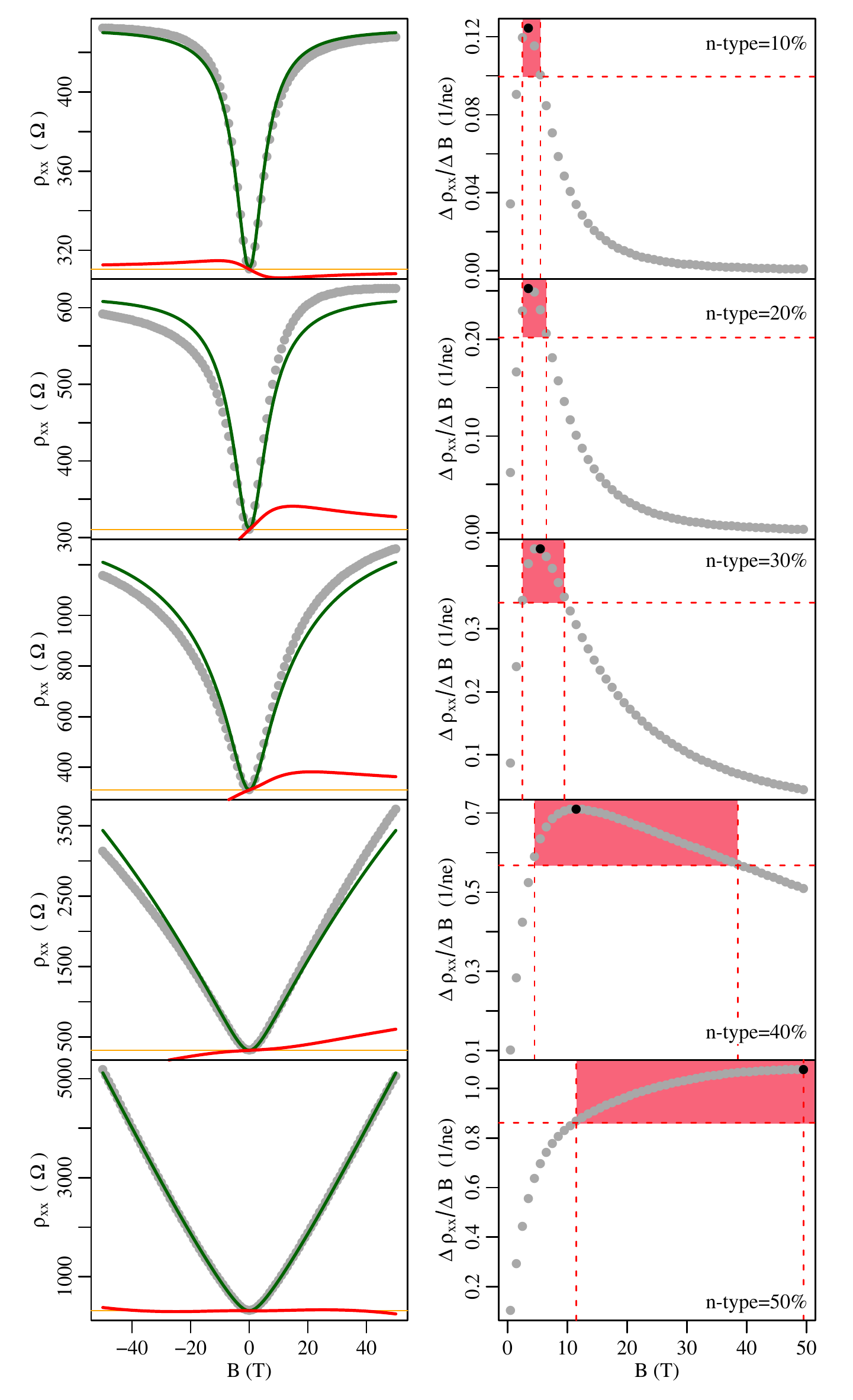}
                        \caption{\textbf{SI} (Color online) The left column of graphs shows the calculation of the symmetric part of $\rho_{xx}$, rhe right column of graphs shows the numerical derivative of $\rho_{xx}$. The percentage of $n$-type tiles is varied from 10$\,$\% at the top graph to 50$\,$\% at the bottom graph. Graphs in the left column: Four-point resistivity $\rho_{xx}$ (grey circles). Voltage difference was calculated at columns 75 and 135 as four-point-probes at the bottom of the network. Green line ist the part of $\rho_{xx}$ symmetric in $B$ and red line is the antisymmetric part in $B$ (with the zero fiel resistivity as offset). Graphs in the right column: Numerical derivative of $\rho_{xx}$ (grey circles). The red shaded area indicates the linear-in-$B$ region of $\rho_{xx}$. It is limited by a decrease of the derivative by $20\,$\% of its maximal value. This condition also sets the field limits, all limits are drawn by red dashed lines.}
    \label{fig:SI4}
\end{figure*}

\begin{figure*}[t!]
    \centering
        \includegraphics[scale=1.0]{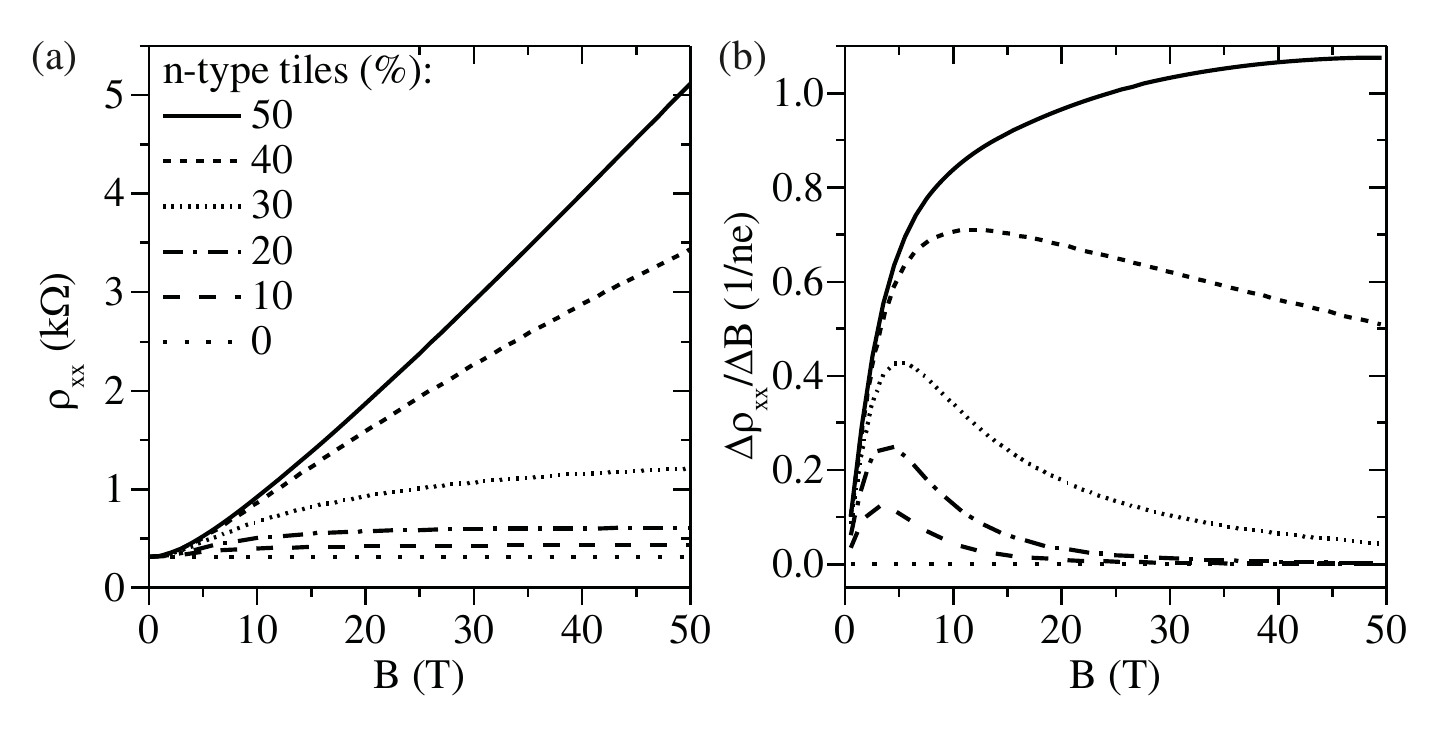}
                        \caption{\textbf{SI} (Color online) Panel (a) $\rho_{xx}(B)$ of Fig. SI \ref{fig:SI3}. (b) Numerical derivative corresponding to curves of $\rho_{xx}$ in (a).}
    \label{fig:SI5}
\end{figure*}

\begin{figure*}[t!]
    \centering
        \includegraphics[scale=1.0]{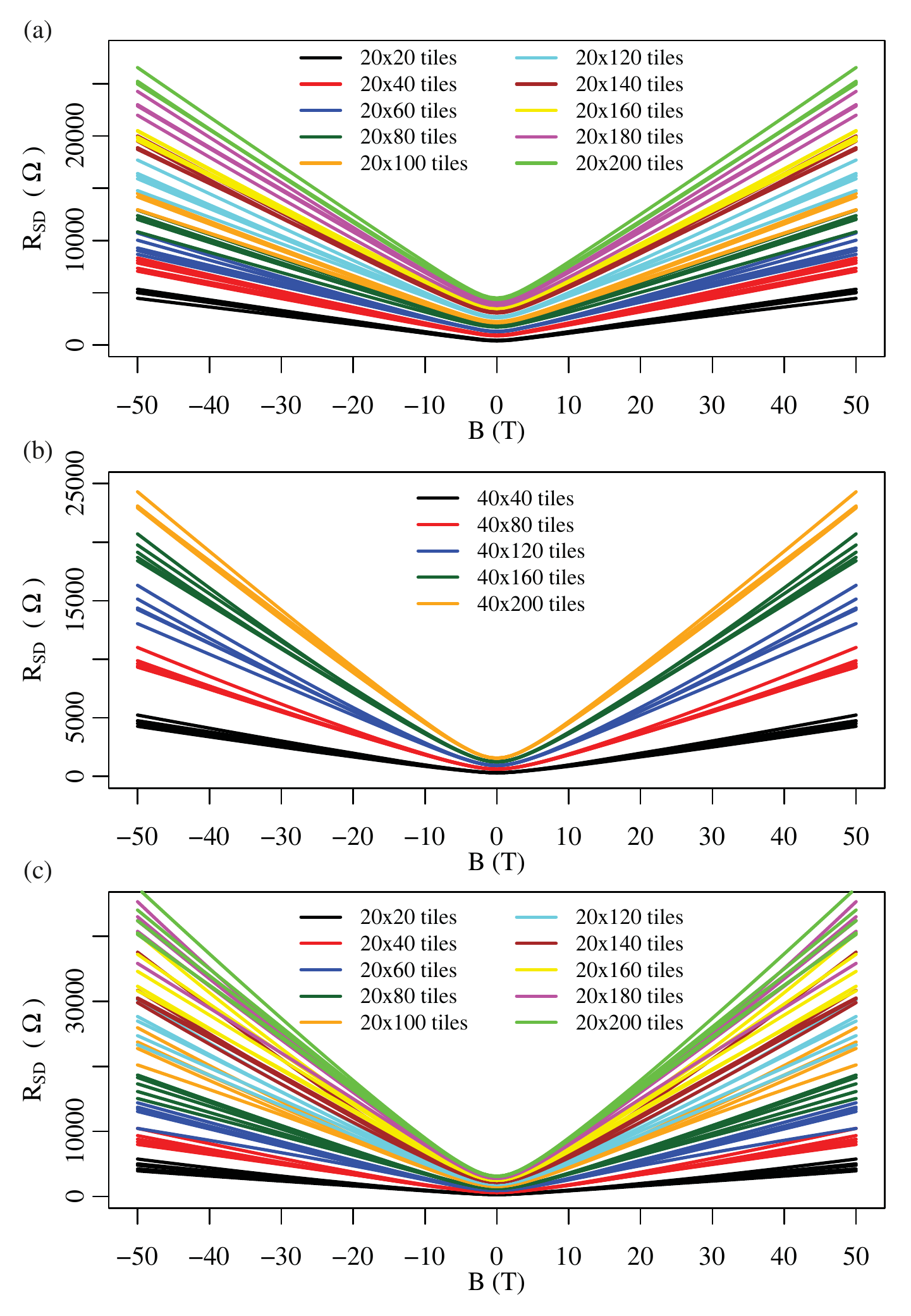}
                        \caption{\textbf{SI} (Color online) $R_{SD}$ for different aspect ratios of the network. Panel (a) corresponding to $n$ disorder (Fig. 4 d). Panel (b) and (c) corresponding to the case of electron hole mixing (Fig. 4 f).}
    \label{fig:SI6}
\end{figure*}

\section{Simulation Parameters}
The standard parameters used for the resistor network simulations are listed in Table SI \ref{tab:SIparam}. The standard parameters correspond to a charge carrier density of $n \approx 6\cdot10^{16}/\mathrm{m}^2$. The parameters deviating from these standard values are always given in the main manuscript. In every case where statistical variations of the parameters for different tiles were made, only one parameter was varied and the others were kept fix. To keep the picture simple a flat distribution with total width $\Delta x$ around a mean value $\left\langle x\right\rangle$ was choosen for the parameter variation.
\begin{table}
\caption{\label{tab:SIparam} Standard parameters used for network simulations.}
\begin{ruledtabular}
\begin{tabular}{llp{0.5\columnwidth}}
Parameter&Value&Comment\\
$\rho$ [$\Omega$]&104&sheet Resistance inside the tile.\\
$\mu$ [m$^2/$Vs]&1&charge Carrier mobility inside the tile.\\
$\delta_t$ [rad]&$0.1$&Opening angle of the terminal.\\
$n_{max}$&1000&Cut of index for approximation of the series appearing in the calculation of $\boldsymbol{z}(\theta_{tile})$.
%\br
\end{tabular}
\end{ruledtabular}
\end{table}

\section{Color Plots}
In the color plots in the main manuscript each tile is represented by one colored element. Black elements denote that the value ist out of the color scale, white elements denote non defined values, such as negative values on a log-scale.

\end{document}